\input{psfig}

\documentstyle[11pt,aaspp4,epsf]{article}
\begin{document}
\slugcomment{Published in {\em Nature} 22 November 2001}

\pagestyle{plain}
\setcounter{page} {1}
\begin{center}
{\Large \bf Gravitational collapse and non-self similarity in the L-T relation 
}
\end{center}
\vskip2cm
\begin{center}
{\bf Antonino Del Popolo$^1$,$^2$, N Hiotelis$^3$ and N. Ercan$^1$}
\\~
\\~
$1$  Bo$\breve{g}azi$\c{c}i University, Physics Department,
     80815 Bebek, Istanbul, Turkey\\
$^2$ Dipartimento di Matematica, Universit\`{a} Statale di Bergamo, Piazza Rosate 2, Bergamo, ITALY

$^3$ 1st Experimental Lyceum of Athens, Ipitou 15, Plaka, 10557, Athens, Greece,

\end{center}

\vskip 2cm

\abstract
We derive the luminosity-temperature relation for clusters of galaxies 
by means of a modification of the self-similar model to take 
account of angular momentum acquisition by protostructures
and of an external pressure term in the virial theorem. 
The fundamental result of the model is that gravitational collapse, which takes account of angular momentum acquisition, can explain the non self-similarity in the L-T relation, in disagreement with the largely accepted assumption that heating/cooling processes and similar are fundamental in the originating the non-self similar behavior (shaping) of the L-T relation.

\vskip 1cm
\section{Introduction}

Till some years ago, the cluster structure was considered to be scale-free, which means that the global properties of clusters, such as halo mass, luminosity-temperature, and X-ray luminosity would scale self-similarly$^{19}$. 

The self-similar model (SSM)$^{19}$, 
is obtained assuming that gas density 
or the baryon number density, $n$, is proportional to the average Dark Matter density, $\rho$, and that the 
virial radius, $R_{\rm vir}$ is proportional to $R_{\rm X}$. 
In this way, one obtains, according to this last, $L \propto M \rho T^{1/2}$. In fact, $T \propto M/R_{\rm vir}$, 
$n \propto \rho \propto M/R_{\rm vir}^3$, $R_{\rm vir} \propto R_{\rm X}$ and $L \propto \int_0^{R_{\rm vir}} \rho^2 T^{1/2}r^2 dr \propto \rho^2 T^{1/2} R_{\rm vir}^3$, and remembering that $R_{\rm vir} \propto (M/\rho)^{1/3}$, leads to $L \propto \rho M T^{1/2}$ or remembering that 
$R_{\rm vir} \propto (T/\rho)^{1/3}$, one gets $L \propto \rho^{1/2} T^2$.  
This last result is inconsistent with observed correlation close to $L \propto T^3$,$^{12}$ 
and a further steepening at the temperature of galaxy groups is indicated for th emission not associated with single galaxies$^{28}$. 


The $L \propto T^3$ relation, has been interpreted as an 
indication that non-gravitational processes
should influence the density structure of a cluster's core, where most
of the luminosity is generated$^{3, 13, 20, 24}$.
One way to obtain a scaling law closer to observational value is to have non-gravitational energy injected into intracluster
medium (ICM) before or during cluster formation. This solution, called pre-heating, was originally invoked to 
solve two related problems: a) to explain
\footnote{Kaiser's self similar model predicts $L \propto T^{3.5}$, while Evrard \& Henry$^{13}$ obtained the relation $L \propto T^{11/4}$} 
the apparent negative evolution of the X-ray cluster luminosity function$^{16,35}$
from the Einstein Medium Sensitivity Survey in a $\Omega_{\rm m}=1$ Universe ; b) to explain 
\footnote{In this case pre-heating was in form of supernovae-driven galactic winds} 
why groups and low-mass clusters seem to have higher X-ray temperatures than expected based on member velocity dispersions$^{36}$, or $1< \beta_{\rm spec} \propto \frac{\sigma^2_{\rm v}}{T_{\rm x}}$.

According to a largely accepted view, the X-ray luminosities of low-temperature clusters are small because their gas is less centrally concentrated than in hotter clusters, an effect that has been attributed to an universal minimum entropy level in intracluster gas, resulting from supernova heating$^{7, 28, 33}$, 
from heating by active nuclei$^{33}$ 
or from radiative cooling$^{4, 27, 33}$. 
In other terms for some reasons the core gas is less high than that expected in the self-similar model. For example, an early episode of uniformly distributed supernova feedback, could rectify the problem by heating the uncondensed gas and therefore making it harder to compress in the core. 
In other words,
the models with pre-heating and similar gives rise to the quoted break
because they change the density in the core. 
During the hierarchical buildup an energy input preheats the gas before it falls into new groups and clusters, so hindering its flow into the latter. The core density shall decrease and so the luminosity.

\end{document}